\documentclass[journal]{IEEEtran}
\IEEEoverridecommandlockouts
\usepackage{cite}
\usepackage{amsmath,amssymb,amsfonts}
\usepackage{algorithmic}
\usepackage{graphicx}
\usepackage{textcomp}
\usepackage{xcolor}
\usepackage{comment}
\usepackage[normalem]{ulem}
\usepackage{subfig}
\usepackage{soul}
\usepackage{hyperref}

\def\BibTeX{{\rm B\kern-.05em{\sc i\kern-.025em b}\kern-.08em
    T\kern-.1667em\lower.7ex\hbox{E}\kern-.125emX}}
\begin{document}

\title{Predicting Socio-Economic Well-being Using Mobile Apps Data: A Case Study of France}
%
%
%

\author{Rahul~Goel,
        Angelo~Furno,
        and~Rajesh~Sharma
\thanks{R. Goel and R. Sharma are with the Institute of Computer Science\\
University of Tartu, Estonia, e-mail: rahul.goel@ut.ee}
\thanks{A. Furno is with LICIT-ECO7 UMR\_T9401, University of Lyon,
ENTPE, University Gustave Eiffel, Lyon, France,
e-mail: angelo.furno@univ-eiffel.fr}
}

%
%

%
\maketitle

\begin{abstract}
Socio-economic indicators provide context for assessing a country's overall condition. These indicators contain information about education, gender, poverty, employment, and other factors. Therefore, reliable and accurate information is critical for social research and government policing. Most data sources available today, such as censuses, have sparse population coverage or are updated infrequently. Nonetheless, alternative data sources, such as call data records (CDR) and mobile app usage, can serve as cost-effective and up-to-date sources for identifying socio-economic indicators. in

This work investigates mobile app data to predict socio-economic features. We present a large-scale study using data that captures the traffic of thousands of mobile applications by approximately 30 million users distributed over 550,000 km$^2$ and served by over 25,000 base stations. The dataset covers the whole France territory and spans more than 2.5 months, starting from 16$^{th}$ March 2019 to 6$^{th}$ June 2019. Using the app usage patterns, our best model can estimate socio-economic indicators (attaining an R-squared score upto 0.66). Furthermore, using models' explainability, we discover that mobile app usage patterns have the potential to reveal socio-economic disparities in IRIS\footnote{IRIS is a region used to divide the country into units of similar population size.}. Insights of this study provide several avenues for future interventions, including user temporal network analysis to understand evolving network patterns and exploration of alternative data sources.

\end{abstract}


\section{Introduction}\label{sec:Intro}
One of the fundamental pillars for the growth of economy is the fast-growing metropolitan areas. However, the prevalence of societal disparity is a major concern that restricts the development of a nation. For example, the increased clustering of individuals based on gender, religion, language, or color in metropolitan cities comes at a tremendous cost in terms of segregation~\cite{goel2021understanding,goel2021studying,goel2021overall}. In a nutshell, segregation is defined as the degree to which two or more demographic groups are separated, and it has drawn much attention for many decades~\cite{farley1994changes,hewstone2005intergroup}. Segregation has far-reaching consequences that go beyond economics. In various parts of the same city, individuals may experience linguistic isolation, widely divergent levels of education, health care, employment, and other resources~\cite{santiago2019framework,ward2018neighborhood,azhdari2018exploring}. The development and successful implementation of policies to address these issues need fine-grained, often updated data on income, education, and inequality across metropolitan regions. However, most data sources used today, such as demographic censuses or surveys, have sparse population coverage or are updated infrequently due to their high costs. As a result, they do not fit the swift evolution that urban societies are experiencing nowadays.

Researchers have used alternate digital traces, such as Mobile call data records and mobile application usage, to understand societal well-being~\cite{ucar2021news,zhou2016general,nanavati2008analyzing}. According to \cite{datareportal}, there were 5.31 billion unique mobile phone users worldwide in January 2022 – equivalent to 67.1 percent of the world's total population. Of these mobile users, 4.95 billion people use the Internet – equal to 62.5 percent of the world's total population. What is more, fewer than 3 billion people worldwide do not use the Internet, marking another important milestone in the universal accessibility of mobile phones. These figures show mobile reach and the significance of call and digital data in studying people worldwide.

\textit{Call} and \textit{Digital} data has recently been recommended as an alternate source for determining socio-economic status \cite{ucar2021news,marquez2017not,marquez2020identifying,singh2019urban}. Researchers have created socio-economic prediction models with unprecedented temporal and geographical resolutions owing to the increasing usage of mobile devices, social media, and the expanding availability of ubiquitous satellite data~\cite{soto2011prediction,gao2019computational,dong2019predicting,blondel2015survey}. Human mobility and social contacts were shown to be linked to higher income in data from mobile phones and social media~\cite{blumenstock2016fighting,pappalardo2016analytical,steele2017mapping,llorente2015social}. Although these methodologies are pretty effective in forecasting socio-economic features in underdeveloped countries, they are only somewhat accurate in developed countries with more nuanced variations in mobile phone adoption~\cite{hashemian2017socioeconomic,jean2016combining,abitbol2020interpretable}.

\textbf{Research question and highlights: }Considering the literature on Call Data Records (CDR)~\cite{goel2021understanding,goel2021studying,goel2021overall} and past research to predict socio-economic features using mobile service consumption data~\cite{ucar2021news}, we show that mobile application usage patterns would be able to predict socio-economic features in a nation. The primary goal of using mobile digital data to anticipate socio-economic features is to eliminate or replace costly and time-consuming censuses. This leads to our research question, that is \textbf{how accurately we can predict socio-economic conditions using mobile applications usage patterns}. The same research question is answered in~\cite{ucar2021news} using mobile application data and population structure 
variables from census data. However, the focus of our research is solely on the use of mobile applications to predict socioeconomic features. The reason for not utilizing population 
data is that this study attempts to provide an alternative to expensive and time-consuming census chores to collect socioeconomic information. The highlights of this study are as follows:
\begin{enumerate}
    \item We conduct a large-scale investigation using nationwide and multi-source datasets. We utilized three different datasets: (i) \textit{Mobile applications usage data} of size 2TB (terabyte) spanning more than 2.5 months by approximately 30 million subscribers distributed over more than 550,000 km$^2$ and served by over 25,000 base stations in a major European country with one of the world's largest economy, France~\cite{data}.; (ii) \textit{Socio-economic features} covering economic status, population, and education information for around 49,000 IRIS$^1$; and (3) \textit{Geographical information} provided by the French Institut national de l’information géographique et forestière (IGN) for around 49,000 IRIS.

    \item Using \textit{Mobile apps data}, we extracted three patterns for each IRIS, i.e., week signature on an hourly basis, Revealed Comparative Advantage (RCA), and Standardized Cumulative Utilization (SCU), explained in Section \ref{sec:iris_mobile_apps_usage} in detail. These three extracted patterns cover a different aspects of information from the dataset. The week signature on an hourly basis aims at summarizing the mobile traffic behavior into a meaningful profile capable of showing the whole week's activity pattern. The RCA is an econometrics index capable of capturing the comparative advantage or disadvantage of the consumption of an application in a given IRIS. Lastly, the SCU index captures the statistics aspect of the dataset.
    \vspace{1mm}
    
    Previous studies of predicting socio-economic features used the econometric index RCA for calculating the relative advantage or disadvantage of mobile app usage in IRIS areas 
    and population data from census~\cite{ucar2021news}. In contrast, this paper examines the typical weekly temporal patterns emerging in the use of mobile apps, which we call week signature, on an hourly basis. We discover that the week signature not only helps to predict socioeconomic features but also has the capability of capturing the entire week behavior (weekdays, weekends, morning, afternoon, evening, and night activity). Details are provided in Section \ref{sec:Exp}.

    \item 
    We experiment with various machine learning based regression models trained on patterns (week signature, RCA, and SCU) extracted from the usage of the mobile applications. However, we report the best regression model in our case, CatBoost~\cite{prokhorenkova2018catboost}. The proposed approach achieves an R-squared score up to 0.66 using the CatBoost regression model to estimate economic, educational, and demographic features (Section \ref{sec:Exp}). The results support the research question of this study that  mobile app data can be used to predict socio-economic features. It is to be noted that the results of this study are not directly comparable with~\cite{ucar2021news} since the authors employed census population data along with RCA as independent variables  to predict socio-economic features.

    \vspace{1mm}

    Furthermore, for CatBoost model explainability, we use the SHAP (Shapley additive explanations) value~\cite{lundberg2017unified}. The SHAP is a cooperative game theory-based strategy for increasing the transparency and interpretability of machine learning models. Using SHAP we display the contribution or relevance of top-20 features for the CatBoost model's prediction.
    
    \item Finally, we compared the two CatBoost model results, in which one of them is trained on the mobile applications usage patterns (see Section \ref{subsec:results&Explainability}) and the other is trained on the actual population information or census of France (see Section \ref{subsec:relevance}) for predicting \textit{Income, Gini index (Income inequality), and people with higher education} in IRIS. We find that mobile applications' usage patterns achieve scores that are at least as good as those obtained with census data.
\end{enumerate}

\textbf{Paper organisation. }
Next, we explain three datasets utilized in this study: mobile app traffic, geographical, and socio-economic features in Section \ref{chapter:datasetDescription}. Then, areal weighted interpolation of IRIS in Section \ref{subsc:areal} and extracted the mobile apps patterns in Section \ref{sec:iris_mobile_apps_usage}. Section \ref{sec:eda} explores the socio-economic disparity from the extracted pattern. We cover the result and explainability in Section \ref{sec:Exp}. Finally, conclusions and limitations are outlined in Section \ref{sec:conclusion}.


\section{Dataset Description}\label{chapter:datasetDescription}
This section describes various datasets which are used in this study. First, we discuss the mobile apps (or applications) data (in Section \ref{sec:MobileAppTraffic}), and next, in Section \ref{sec:SocioEconomic} we explain in detail the geographical information and socio-economic features dataset.

\subsection{Mobile Applications Traffic Data}\label{sec:MobileAppTraffic}
This study uses mobile applications dataset which records the uplink (data transferred by the user's mobile) and downlink (data flowing to the user's mobile) byte counts per mobile app at a temporal granularity of 1 minute. The dataset is aggregated for antennas located at base stations (BS). Orange France collected the data which spans for more than 2.5 months, from 16$^{th}$ March 2019 to 6$^{th}$ June 2019. The dataset contains the utilization of multiple mobile apps and covers approximately 30 million subscribers distributed over more than 550,000 km$^2$ and by 25,000 base stations. The data is timestamped at the 1-minute granularity and contains BS latitude and longitude information records for various mobile apps, such as WhatsApp, Twitter, and YouTube- including apps from both Apple Store for ios devices and Google Play for android devices.


\textbf{Mobile applications categorization. }We manually categorized the top 99 percentile mobile applications in terms of traffic load into 19 classes. 
We chose the top 99 percentile mobile apps since only a small number of apps generate a significant volume of data according to the power law~\cite{marquez2017not,shafiq2011characterizing}. The idea of categorizing mobile apps is also explored in past research~\cite{liu2016macro}. The mobile apps that are selected include heterogeneous apps and encompass \textit{Advertising} (e.g., Web Advertising), \textit{Android download} (e.g., Google Play Store), \textit{Apple cloud} (e.g., iCloud storage), etc. Table S1 in the \href{https://drive.google.com/drive/folders/1CvsAuMG15L2Xk71XXkqLDT0DWlxHVr9l?usp=sharing}{supplementary document} contains the detailed information about application categorization.

\textbf{Ethical concern. }Our research follows strong ethical standards. Orange, France provided the data and was responsible for data collection, processing, and storage. They follow the rules of the European Commission's General Data Protection Regulation (GDPR). All these privacy-related activities were also monitored by the Orange Data Privacy Officer and the \textit{CNIL}, the French governmental body responsible for safeguarding privacy in the use of personal data. In this study, we used aggregated data among thousands of people at the antenna level, which does not pose a privacy threat and does not qualify as personal data.

\subsection{Geographical data and socio-economic features}\label{sec:SocioEconomic}
This section describes the geographical and socio-economic features dataset used in this paper. Please note that the mobile application data used in this study is from 2019, whereas the geographical and socio-economic features datasets are from 2018. 
The geographical and socio-economic features datasets from 2018 are chosen for two reasons. First, keeping the geographic information and socio-economic feature consistent is vital. Second, these publicly available datasets are the ones that are temporally closest to our mobile app traffic data.

\noindent\textbf{Geographical data. }We used publicly available geographical information provided by the French Institut national de l’information géographique et forestière (IGN). For the geographical description, we used IRIS-level information. We downloaded the \textbf{Contours IRIS édition 2018} dataset\footnote{https://www.data.gouv.fr/en/datasets/contours-iris-insee-ign/}, which contains coverage area for each IRIS zone in France, and also the IRIS code, name, and type. According to the Institut national de la statistique et des études économiques or National Institute of Statistics and Economic Studies (INSEE) in France, IRIS divides the country into units of equal size in terms of population, also known as IRIS2000. Here, IRIS refers to \emph{aggregated units for statistical information}, and 2000 refers to the target size of 2000 residents per basic unit.

\noindent\textbf{Socio-Economic features.} For understanding the population structure and socio-economic features in France, we leveraged the following datasets from INSEE:

\begin{enumerate}
    \item \textbf{Economic status: }For economic status, we downloaded the \textbf{Revenus, pauvreté et niveau de vie en 2018 (Iris)}\footnote{https://www.insee.fr/fr/statistiques/5055909} dataset~\cite{insee_1}, which contains income information for residential IRIS zones. Due to privacy reasons, IGN does not share features for areas with less than 1000 people. After filtering out IRIS zones with missing values, we have the economic status for 9,145 IRIS zones (out of 48,931) of France as shown in Fig. S1(a) in the \href{https://drive.google.com/drive/folders/1CvsAuMG15L2Xk71XXkqLDT0DWlxHVr9l?usp=sharing}{supplementary document}.  
    
    We work on disposable income, meaning people declare their total incomes once a year (salary, allowances, rent, financial products, etc.) and pay taxes on their total incomes. The disposable income is what they keep after redistribution. Next, we select three features: Poverty, Median income, and Gini Index from the economic statuses, which are vital for understanding the economic situation of the IRIS. The selected features and their definition are shown in Table \ref{table:se_data}.

    \item \textbf{Population information: }For the population information, we utilized the \textbf{Population en 2018}\footnote{https://www.insee.fr/fr/statistiques/5650720} dataset, which contains the population information for various age groups, gender, immigrants, and professional category. The selected features from the population information based on correlation are shown in Table \ref{table:se_data}. For population information, we have data for 45,508 IRIS zones (out of 48,931) as shown in Fig. S1(b) in the \href{https://drive.google.com/drive/folders/1CvsAuMG15L2Xk71XXkqLDT0DWlxHVr9l?usp=sharing}{supplementary document}.

    \item \textbf{Education information: }For education statistics, we used the \textbf{Diplômes - Formation en 2018}\footnote{https://www.insee.fr/fr/statistiques/5650712} dataset, which contains the academic level information (BEPC, BAC, SUP, CAPBEP, etc.). The selected features from the education information are shown in Table \ref{table:se_data}. Similar to population, we have data for 45,508 IRIS zones (out of 48,931) for education information.

\end{enumerate}

\begin{table*}
\centering
\begin{tabular}{ll} 
\hline
\textbf{Socio-Economic Indicator} & \textbf{Definition}                                                                                                          \\ 
\hline
\textbf{Economic conditions}      &                                                                                                                              \\
Poverty                           & Poverty rate at the threshold of 60\% of disposable income per metropolitian median per person (\%)                          \\
Median Income                     & Median income per person (in euros)                                                                                          \\
Gini Index                        & The Gini Index summarizes the dispersion of income across the entire income distribution i.e, it measure income inequality \\ 
\hline
\textbf{Population information}               &                                                                                                                              \\
Total population                  & Total number of people                                                                                                       \\
Pop 0-14                          & Number of people aged between 0 to 14 years                                                                                          \\
Pop 15-29                         & Number of people aged between 15 to 29 years                                                                                         \\
Pop 30-44                         & Number of people aged between 30 to 44 years                                                                                         \\
Pop 45-59                         & Number of people aged between 45 to 59 years                                                                                         \\
Pop 60-74                         & Number of people aged between 60 to 74 years                                                                                         \\
Pop 75+                           & Number of people aged 75 years or more                                                                                               \\
Immigrants                        & Number of people who are immigrants                                                                                                  \\
CS1                               & Number of people aged 15 or more who are Farmer operators                                                                            \\
CS2                               & Number of people aged 15 or more who are Craftsmen, Traders, Company managers                                                        \\
CS3                               & Number of people aged 15 or more who are Managers and higher intellectual professions                                                \\
CS4                               & Number of people aged 15 or more who are Intermediate professions                                                                    \\
CS5                               & Number of people aged 15 or more who are Employees                                                                                   \\
CS6                               & Number of people aged 15 or more who are Worker                                                                                      \\
CS7                               & Number of people aged 15 or more who are Retired                                                                                     \\
CS8                               & Number of people aged 15 or more who fall in Others w/o professional activity                                                        \\
Male                              & Number of people who are male                                                                                                        \\
Female                            & Number of people who are female                                                                                                      \\ 
\hline
\textbf{Education information}                &                                                                                                                              \\
No diploma                        & Number of out-of-school people aged 15 or over with no diploma or atleast a CEP                                                      \\
BEPC or CAPBEP                    & Number of out-of-school people aged 15 or over holding a BEPC, college certificate, DNB, CAP, BEP or equivalent;                     \\
BAC                               & Number of out-of-school people aged 15 or over with a Baccalaureate, professional certificate or equivalent                          \\
SUP                               & Number of out-of-school people aged 15 or over with a higher education diploma at Bac +2 level or more                               \\
\hline
\end{tabular}
\caption{Selected socio-economic features (per IRIS area) with their definitions from INSEE.}\label{table:se_data}
\end{table*}

\begin{figure*}[ht!]
  \centering
  \includegraphics[width=1.6\columnwidth]{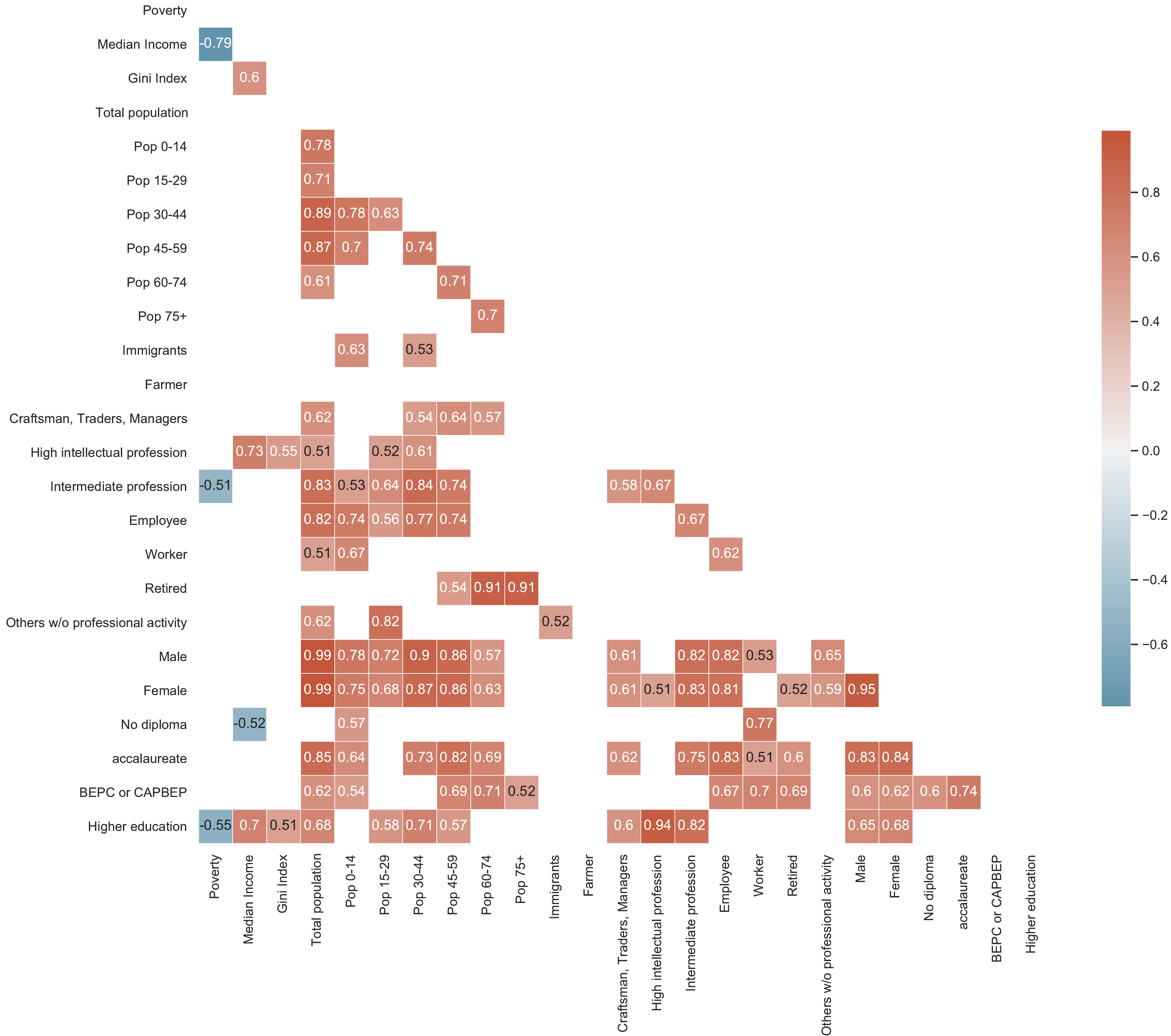}
  \caption{Correlation plot for socio-economic features.}
\label{fig:corr_se}
\end{figure*}

Next, we performed the Pearson's correlation test to study the correlation between all 25 socio-economic features as shown in Figure~\ref{fig:corr_se}. All weak correlation values (between -0.5 and 0.5) are set to null for ease of interpretation. From Figure ~\ref{fig:corr_se}, we observe that poverty is negatively correlated with median income, which is obvious. The median income is positively correlated with high intellectual profession and education, implying that high professionals and highly educated individuals earn more. Similarly, we can observe that individuals with age above 60 years are retired (correlation value as 0.91).

As mentioned earlier, the Geographical data and socio-economic features mentioned in this section are at the IRIS level. However, the mobile app data is at the antenna level. Therefore, we convert the mobile app data to IRIS level using the areal weighted interpolation method.


\section{Areal weighted interpolation of IRIS}\label{subsc:areal}
In this section, we identified the areal weighted interpolation of IRIS using base station location and IRIS geographic information. As mentioned earlier, our mobile app data is aggregated by antennas. These antennas are located at base stations, which means a base station (BS) can correspond to single or multiple antennas. Therefore, as a first step, we mapped all the antennas to their respective BS based on their latitude (LAT) and longitude (LON) information. Next, we followed the technique mentioned in \cite{aasa2021spatial, ucar2021news} for the areal weighted interpolation of IRIS. We begin by modeling the coverage area of each BS using the Voronoi polygon.

\begin{figure*}[ht!]
  \centering
  \subfloat[Voronoi Polygon]{\label{fig:voronoi_polygon}\includegraphics[width=0.5\columnwidth]{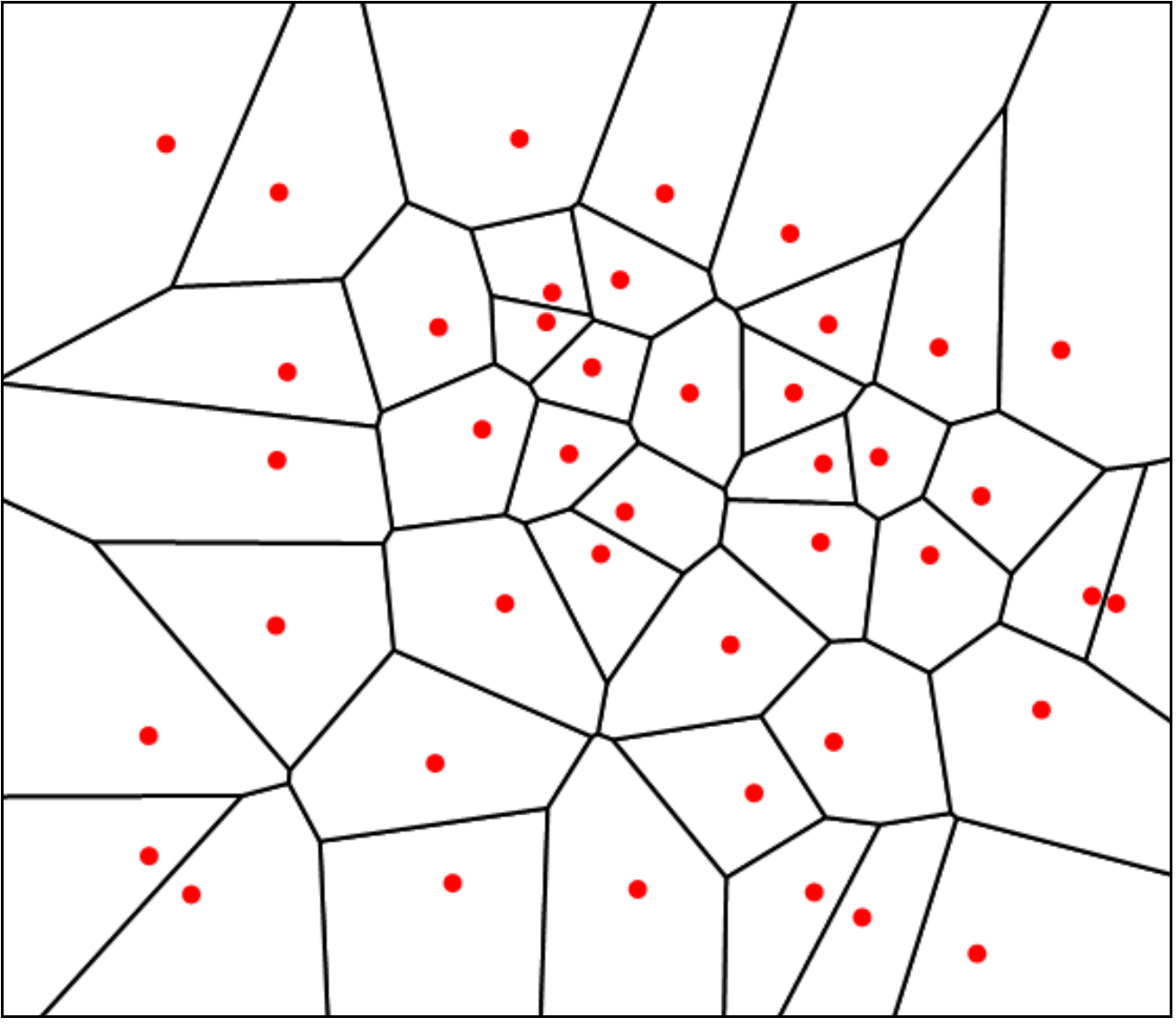}}\hspace{1mm}
  \subfloat[IRIS Zone]{\label{fig:IRIS_zone}\includegraphics[width=0.5\columnwidth]{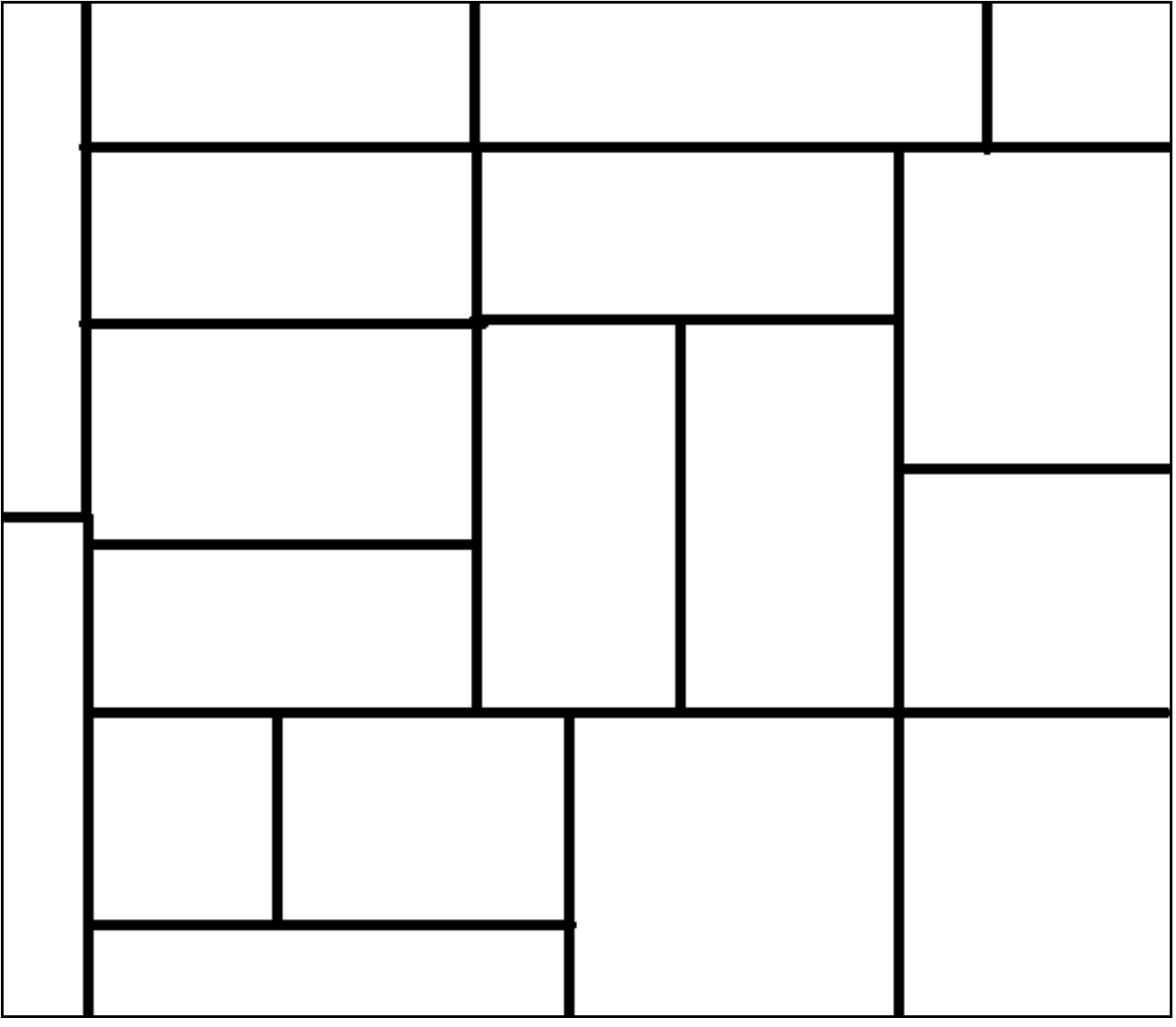}}\hspace{1mm}
  \subfloat[Voronoi and IRIS zone mapping]{\label{fig:voronoi_IRIS_mapping}\includegraphics[width=0.5\columnwidth]{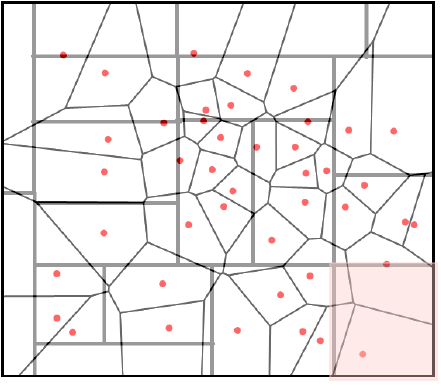}}
  \caption{Areal weighted interpolation for mapping base station (BS) to IRIS. Figure \ref{fig:voronoi_polygon} shows the Voronoi polygon generated using BS in a bounded plane. The IRIS zone area coverage is shown in Figure \ref{fig:IRIS_zone}. In Figure  \ref{fig:voronoi_IRIS_mapping}, the areal weighted interpolation of mobile app usage into the IRIS zone is shown. The highlighted box at the bottom-right of Figure \ref{fig:voronoi_IRIS_mapping} shows an IRIS, comprising the weighted data from four BS. (best seen in color)}
\label{fig:areal_interpolation}
\end{figure*}

Let us suppose we have finite number of BS, $N$, in the two-dimensional Euclidean plane, and assume that 2 $\leq$ $N$ $< \infty$. The $N$ number of BS are labeled by $bs_1$, $bs_2$,..., $bs_N$ with the Cartesian coordinates ($x_{1}$, $y_{1}$), ($x_{2}$, $y_{2}$), ..., ($x_{N}$, $y_{N}$) or location vectors $X_1$, $X_2$, ..., $X_N$. The $N$ points are distinct in the sense that $X_i \neq X_j$ for $i \neq j$, where $i$, $j$ $\in I_N$  = \{1, · · · , $N$\}, where $I_N$ represents a set of $N$ natural numbers. Let $bs$ be an arbitrary BS in the Euclidean plane with coordinates ($x$, $y$) or location vector $X$. Then the Euclidean distance from $bs$ to $bs_i$ is given by

\begin{equation}
    d(bs, bs_i) = ||X - X_i|| = \sqrt{(x - x_i)^2 + (y - y_i)^2}
\end{equation}

If $bs_i$ is the nearest point from $bs$, we have the relation $||X - X_i|| \leq ||X - X_j||$ for $j \neq i$, where $j \in I_N$. Let B = \{$bs_1$, $bs_2$,..., $bs_N$\}, then the Voronoi polygon region associated with $bs_i$ is given by

\begin{equation}
    V(bs_i) = \{X\ s.t.\ ||X - X_i|| \leq ||X - X_j||\ for\ j \neq i, j \in I_N\}
\end{equation}

and the Voronoi diagram generated by $B$ is given by the set

\begin{equation}
\label{eqn:voronoiDiagram}
    V = \{V(bs_1), V(bs_2), ..., V(bs_N)\}
\end{equation}

Figure \ref{fig:voronoi_polygon} provides an example of the Voronoi polygon generated using BS in a bounded plane. In equation \ref{eqn:voronoiDiagram}, we have defined a Voronoi diagram in an unbounded plane. However, in most cases we deal with a bounded region. Now, the bounded Voronoi diagram generated by $B$ for a bounded region $R$ is given by the set

\begin{equation}
    V_{bnd} = V \cap R = \{V(bs_1)\cap R, ..., V(bs_N)\cap R\}
\end{equation}


Let suppose we have finite number of IRIS zones, $Z$, which are labelled as $I_1,I_2,...,I_Z$. Then, the areal weight for IRIS zone $i$ with respect to all the BS bounded Voronoi polygon is given by

\begin{equation}
    W(I_i) = \left\{\frac{A(V_{bnd}(bs_1) \cap I_i)}{A(V_{bnd}(bs_1))}, ..., \frac{A(V_{bnd}(bs_N) \cap I_i)}{A(V_{bnd}(bs_N))} \right\}
\end{equation}

where, $V_{bnd}(bs_i)$ represents the bounded Voronoi polygon for $bs_1$, $(V_{bnd}(bs_1) \cap I_i)$ represents the overlapping area of Voronoi polygon for $bs_1$ and IRIS zone $i$. A($\cdot$) represents the area of the region or polygon within the parenthesis. Hence, $\frac{A(V_{bnd}(bs_1) \cap I_i)}{A(V_{bnd}(bs_1))}$ is the areal weight for IRIS zone $i$ with respect to the $bs_1$ bounded Voronoi polygon. This areal weight is further multiplied by the total traffic recorded for the respective BS to calculate the traffic consumption. 

Next, the identified aerial weights are multiplied by the total traffic recorded for an app category on the respective BS of an IRIS to calculate the traffic consumption by an app category in the IRIS zone. For example, let us consider the generic IRIS zone $Z$, containing three BS with aerial weights $W_1, W_2$ and $W_3$ and traffic $T_1, T_2$ and $T_3$ for app category $A$ at any given time. The total traffic consumption for app category $A$ in the considered IRIS $Z$ will amount to $T_1*W_1 + T_2*W_2 + T_3*W_3$. After this, we have per-application traffic data entries recorded as the uplink and downlink at a 1-minute granularity and aggregated by IRIS. As a next step, we also sum up the uplink and downlink byte counts, which results in the dataset with per-application traffic data entries recorded as the total byte counts at a temporal granularity of 1 minute and aggregated by IRIS. In the next section, we refer to this dataset as $\mathcal{D}$.



\section{Mobile Apps Usage Features for IRIS sectors}\label{sec:iris_mobile_apps_usage}
\label{sec:features}
Predicting socio-economic features using mobile app data significantly depends upon distinctive usage patterns in mobile phone application data. In this section, we derived a set of metrics from mobile phone app data able to capture useful information from the large-scale raw mobile phone app data to be later used in the classification stage. In particular, we identified 
week signature (Section \ref{subsec:weekly_signature}), Revealed Comparative Advantage (RCA)~\cite{balassa1965trade} utilization (Section \ref{subsec:rca_pattern}), 
and standardized cumulative utilization (Section \ref{subsec:scu_pattern}), 
formally defined in the following subsections.

\subsection{Typical Week Signature (TWS) for IRIS}\label{subsec:weekly_signature}
Let us consider our mobile app consumption dataset $\mathcal{D}$, describing the usage of a set of mobile apps \emph{A} for the \textit{Orange} subscriber population during a set of \emph{d} days. For the definition of the signature, let us define the signature support $\Delta$ of one week, i.e., from Monday to Sunday, $\Delta$ = \{mon, tue, wed, thu, fri, sat, sun\}. Let then us denote as \emph{d$^\delta\subset$} \emph{d} the set of days in the dataset \emph{D} that correspond to the day of the week $\delta \in \Delta$, with
$\bigcup_{\delta\in \Delta}\emph{d}^{\delta}$ = \emph{d}. Then, the generic element in the signature of an IRIS \emph{i} for mobile app \emph{a} at time \emph{t} for day of the week $\delta$ is:

\begin{equation}
s_i(\delta,t,a) = M(\{v_i(d,t,a)|d\in d^{\delta}\}), \hspace{3mm}\forall i\in \emph{I}, a\in \emph{A},
\label{eq:signature}
\end{equation}

\noindent where, $v_i(d,t,a)$ describes the total mobile app \emph{a} usage within the IRIS \emph{i} at time slot (hour in our case) \emph{t} of day \emph{d} from dataset $\mathcal{D}$. M($\cdot$) represents the median of the set within parenthesis. Also in this case, $\delta$ is small with respect to the overall set of \emph{d} days, which implies data compression. \emph{I} and \emph{A} represents set of IRIS and applications respectively.

Signatures then undergo a standard normalization. To that end, each element obtained in Eq.~\ref{eq:signature} is normalized with respect to the mean and standard deviation of all elements referring to the same IRIS and mobile application. Formally, for a generic element of IRIS \emph{i} and mobile application \emph{a} is

\begin{equation}
    \tilde{s}_i(\delta,t,a) = \frac{s_i(\delta,t,a) - \mu(s_i(a))}{\sigma(s_i(a))}, \hspace{1mm}\forall \delta \in \delta, i\in \emph{I}, a\in \emph{A},
\end{equation}

\noindent where $\mu(s_i(a))$ and $\sigma(s_i(a))$ denote the mean and standard deviation of the set of elements concatenated in the signature \emph{s}$_i$.

The techniques for the construction of a TWS for per-application traffic data, at a temporal granularity of 1 hour and for each IRIS~\cite{furno2016tale,furno2017joint} requires to process the dataset $\mathcal{D}$ through three phases. These phases aim at summarizing the mobile traffic activity in each unit areas into a meaningful profile, i.e., the IRIS signature for the mobile application category. The steps involved in generating the TWS are as follows:

\begin{enumerate}
    \item We converted the timestamp value to the corresponding ``\textit{Hour within week}'' value, which means that all the timestamp values represent the hour and day of the week. For example, all the timestamp values on Monday between 00:00 to 00:59 hours, will become 1 irrespective of the date. Similarly, all the timestamp values on Tuesday between 16:00 to 16:59 hours, will become 41 (24+17) irrespective of the date.
    \item Next, we aggregate the data on ``\textit{Hour within week}'' value, and calculate the median total byte count per-application for each IRIS. We considered the median value as denoising component and extracts information deemed to be representative of the typical mobile traffic activity in an IRIS, isolating it from the inherent noise in the data. 
    \item  Lastly, we standardised the signature making it independent from the absolute volume of mobile traffic recorded at an IRIS. This allows comparing the per-application activity at IRIS level on the sole basis of the traffic data variations. This we named as \textit{Typical Week Signature}.
\end{enumerate}

The \textit{TWS} is a relevant feature from mobile phone app usage data due to its capability to depict the relative dynamics of mobile app consumption over the course of an entire week under the premise of cyclic (weekly) regularity at the scale of an IRIS region.

\subsection{IRIS Revealed Comparative Advantage (RCA) utilization}\label{subsec:rca_pattern}
Depending on the nature of the data, different mobile apps produce a heterogeneous traffic volume over the network. For example, videos generate a significantly larger load per session than messaging. As a result, the traffic volume across app categories might vary by several orders of magnitude. Therefore, we also computed the revealed comparative advantage (RCA)~\cite{balassa1965trade}, a relative measure that can compare traffic across IRIS zones and app categories. The RCA index for IRIS zone $z$ and app category $a$ is calculated as follows:



\begin{equation}
RCA_{za} = \frac {T_{za}/\sum_{a' \in A} T_{za'}} {\sum_{z'\in Z} T_{z'a}/\sum_{z'\in Z, a'\in A}T_{z'a'}}
\end{equation}

where $T_{za}$ is the median hourly traffic per person in IRIS zone $z$ for apps category $a$; $\sum_{a' \in A} T_{za'}$ is the median hourly traffic per person in zone $z$ combinedly generated by all considered apps; $\sum_{z'\in Z} T_{z'a}$ is the median hourly traffic per person generated by app $a$ in all zones; $\sum_{z'\in Z, a'\in A}T_{z'a'}$ is the median hourly traffic per person, aggregated over all zones and apps. A value of $RCA_{za}$ $>$ 1 indicates more than average usage of app $a$ in zone $z$; on the other hand, $RCA_{za}$ $<$ 1 indicates lower than average usage of app $a$ in zone $z$. 


\subsection{IRIS standardized cumulative utilization}\label{subsec:scu_pattern}
Next, we extracted the cumulative utilization of apps in different IRIS. This is defined as the total data transmitted in byte counts for each app aggregated for the IRIS. This is instead a simple metric compared to the \textit{week signature} and \textit{RCA}. The motivation behind calculating this information is to check the effectiveness of straightforward metrics to predict socio-economic features. We take the total data transmitted bytes during the 2.5-month observation period for each IRIS zone and app category. Finally, we calculate the standardized cumulative utilization (SCU) as follows:

\begin{equation}
SCU_{za} = \frac{C_{za} - mean(\forall_Z C_{Za})}{sd(\forall_Z C_{Za})}
\end{equation}

\noindent where $C_{za}$ is the cumulative traffic in IRIS zone $z$ for app category $a$; $mean(\forall_Z C_{za})$ is the mean total traffic in all zones $Z$ generated by app $a$; $sd(\forall_Z C_{za})$ is the standard deviation of total traffic in all zones $Z$ generated by app $a$. Therefore, the SCU value quantifies how far off the mean traffic generated by a mobile app in a given IRIS zone is in terms of standard deviations.

\section{Social Disparity From Apps Usage Pattern}\label{sec:eda}
In this section, we explore the potential of all three previously extracted mobile app usage features (i.e., weekly signature, RCA index, and SCU values, introduced in Section~\ref{sec:features}) as predictors of the socio-economic characteristics of IRIS sectors. In particular, we look for a significant disparity in mobile phone app usage within groups of IRIS sectors with similar socio-economic features (discussed in Sec.~\ref{sec:SocioEconomic}). This analysis represents a preliminary exploratory step toward predicting socio-economic sector features using mobile phone app consumption. To do so, we considered the median income as a use case. For better visualization, we used the quantile method to classify IRIS based on median income into three classes: Low, Medium, and High, with an equal number of IRIS in each class. Three classifications are created in order to demonstrate socio-economic features' distinct differences if any exist.

\subsection{Week Signature}\label{subsec:eda_weekly_signature}
In Section \ref{subsec:weekly_signature}, we discussed the procedure to extract TWS with one-hour granularity for each application category. Here, we show that the TWS helps predict socio-economic features. As we know, the TWS for an application in an IRIS is a vector with 168 values (7 days multiplied by 24 hours, i.e., $24\times7=168$). As a first step, we generate TWS for each income class (Low, Medium, and High). To do so, we compute the mean TWS of all IRIS in a class. 

\begin{figure*}[ht!]
  \centering
  \subfloat[Week signature. Here, x-axis represents hour within week and y-axis shows the mean TWS value of the income class.]{\label{fig:eda_sign}\includegraphics[width=2\columnwidth]{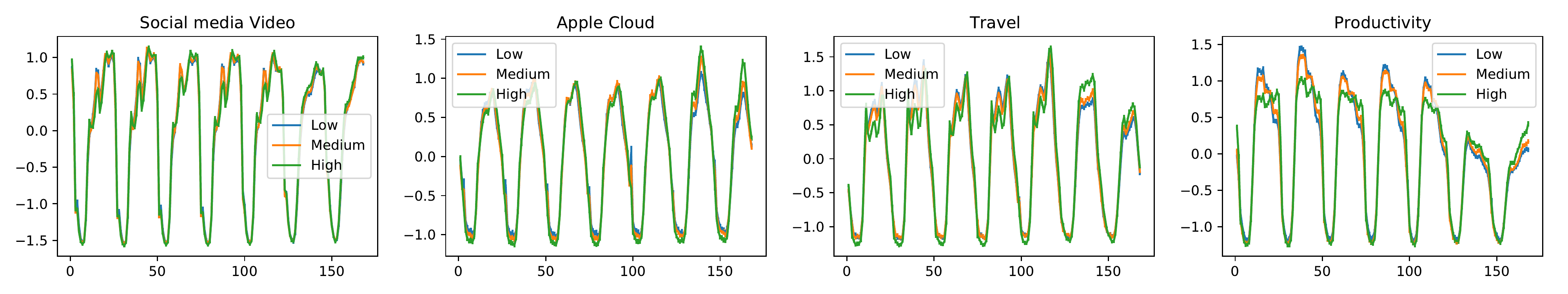}}\\
  \subfloat[RCA index. Here, x-axis represents application category and y-axis shows the mean RCA index of the income class.]{\label{fig:eda_rca}\includegraphics[width=2\columnwidth]{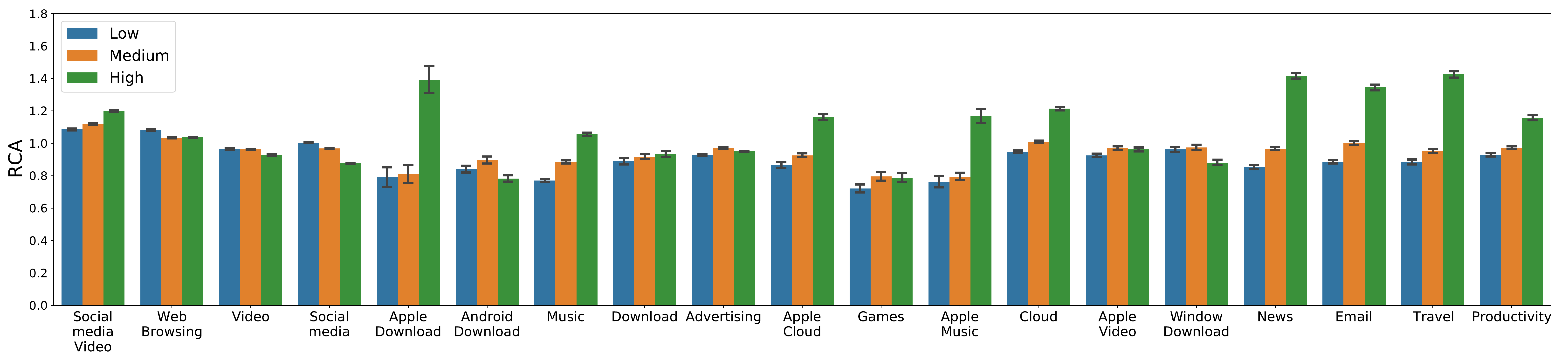}}\\
  \subfloat[SCU index. Here, x-axis represents application category and y-axis shows the mean SCU index of the income class.]{\label{fig:eda_scu}\includegraphics[width=2\columnwidth]{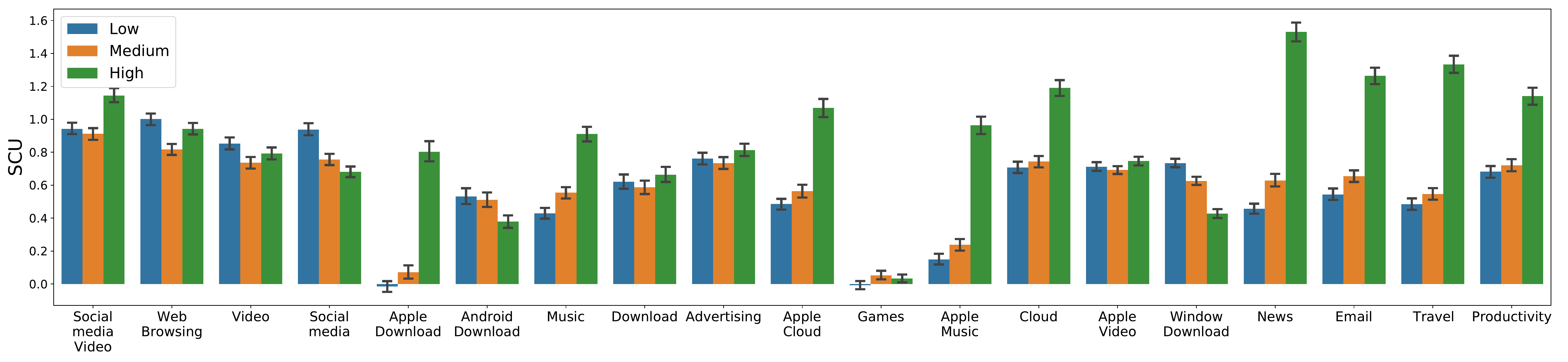}}
  \caption{IRIS income class (Low, Medium, and High) disparity from applications usage pattern. Here, Figure \ref{fig:eda_sign} shows the week signature patterns for \textit{Social media Video, Apple Cloud, Travel}, and \textit{Productivity} apps. Figure \ref{fig:eda_rca}, and \ref{fig:eda_scu} shows RCA and SCU index respectively with the 95\% confidence interval. Furthermore, the line or bar with blue, orange, and green color represents the Low, Medium, and High income IRIS. (best seen in color)}
\label{fig:EDA_RCA_SCU}
\end{figure*}

Figure \ref{fig:eda_sign} shows the mean TWS for each income class. Here, we included the outcomes of four application categories: Social media video, Apple Cloud, Travel, and Productivity. These app categories are included due to their distinct usage pattern for income class. The x-axis depicts the hour within a week, and the y-axis represents the mean TWS of all IRIS in an income class. The blue, orange and green lines reflect the Low, Medium, and High-income IRIS, respectively. 
It is to be noted that the TWS are standardized, therefore any choices we make moving forward in this section should be interpreted as relevant to the usage as a whole (which is subtracted from the signal). In other words, when we use the word ``more", it does not mean ``more" in the volume sense but rather "more" in a relative sense, i.e., in relation to the average standardized weekly usage for a particular income category.

In the social media video, we can observe that the usage pattern for all three classes is consistent, regardless of whether it is a weekend or weekday. However, there is a difference in the use pattern of High-income IRIS compared to Low and Medium income IRIS during the weekday late morning, where Low and Medium income IRIS use social media video more than High-income IRIS. This is most likely because the high-income IRIS includes company managers and highly intellectual professionals, as shown in Figure \ref{fig:corr_se}, so they spend their morning schedule working instead of watching social media videos. The social media video usage is even more intriguing because the patterns of app utilization do not alter much between weekdays and weekends. However, the apparent difference in weekend early morning social media video usage among different income classes demonstrates the significance of TWS as a good predictor. 

Concerning the Apple Cloud app feature, we observe that all three income classes have identical consumption patterns during the weekdays. However, the usage is relatively more distinct on weekends than on weekdays. Furthermore, the High-income group generates the most data on average on the Apple Cloud app, followed by the Medium, and then the Low-income groups. 
When Apple devices are plugged in, locked, and connected to Wi-Fi, Apple cloud (iCloud) performs a daily backup automatically. When most people get the weekend off from work, these conditions are more likely to be met. As a result, the high-income group's weekend usage of the Apple cloud suggests that they own more Apple devices than individuals of the low and medium-income groups.

In Travel, usage is more on Friday. This suggests that irrespective of income group, most individuals use traveling-related mobile applications on Friday. This usage is probably because more people travel on Friday after work to spend the weekend with their family. On weekdays, the usage patterns of the Low and Middle-income classes are comparable, while the usage of the High-income class is less. However, on weekends, this usage pattern is reversed. This suggests that Low- and medium-income individuals travel less on weekends than high-income groups. In the Productivity, usage order is Low$>$Medium$>$High on weekdays. However, high-income group usage is more in the late evening. On the weekend, this order is reversed, which suggests that the high-income group, which includes company managers, and high intellectual professions use productivity applications more than low and medium-income groups on weekends. Another intriguing finding we can infer is that Tuesday is the busiest day of the week when productivity app consumption is at its peak for all the income groups. This further support our hypothesis that some application categories play a more vital role in identifying socio-economic features than others. From these observations, we can conclude that the TWS helps predict socio-economic features.

\subsection{RCA Index}\label{subsec:eda_rca_index}
In Section \ref{subsec:rca_pattern}, we discussed in detail the extraction of RCA index. This index indicates that if $RCA_{za}>1$, the IRIS $z$ utilizes app category $a$ more than the other IRIS. If RCA is less than unity, the IRIS is said to have less app consumption than other IRIS. Therefore, it reveals the comparative advantage or disadvantage of the consumption of an application in an IRIS.

Figure \ref{fig:eda_rca} depicts the mean RCA index value for the three income classes across all application categories. The x-axis represents the various app categories, while the y-axis indicates the mean RCA index. The low, medium, and high-income IRIS are represented by the blue, orange, and green bars, respectively. The 95\% confidence interval for each income class and app category is also included in the bars. According to the figure, high-income IRIS utilizes more the app categories like Apple Download, Music, Apple Cloud, Apple Music, Cloud, News, Email, Travel, and Productivity. Furthermore, middle-income IRIS prefers Android and Windows downloads. Finally, low-income IRIS are more likely to use web browsing, video, and social media. These findings align with the previous research, e.g., low-income families are more likely to utilize online social media than higher-income families~\cite{lenhart2010social,nasirudeen2017impact}.

\subsection{SCU Values}\label{subsec:eda_scu_values}
In Section \ref{subsec:scu_pattern}, we discussed in detail the extraction of SCU values. The value of the $SCU_{za}$ tells us how many standard deviations the value is away from the mean consumption among all the IRIS of the application category $a$. If an $SCU_{za}$ is equal to 0, it is on the mean. A positive $SCU_{za}$ value indicates the value is higher than the mean consumption among all the IRIS for application category $a$. For example, if an $SCU$ value equals +1, it is one standard deviation above the mean consumption.

Figure \ref{fig:eda_scu} shows the mean value of the SCU index of the three income classes for all application categories. The x-axis represents the various app category, and the y-axis represents the mean SCU index. The blue, orange, and green bars show the Low, Medium, and High-income IRIS respectively. The bars also have a 95\% confidence interval for each income class and app category. The figure shows that app categories such as Social media Video, Apple Download, Music, Apple Cloud, Apple Music, Cloud, News, Email, Travel, and Productivity are utilized more by high-income IRIS. The plausible explanation of the observed differences in app usage in terms of socio-economic disparity can be two-fold. First, wealthier individuals who work and travel more often than low- and middle-income groups on weekends use Apple services more frequently. Second, those with high incomes consume a significant amount of music and videos. This is somewhat similar to what we observed with the RCA index. However, these two index captures very different aspect of data utilization. Then, there is Web Browsing, Video, Social media, Android Download, and Window Download, which low-income IRIS uses more. This implies that low- and middle-income groups have a higher prevalence of Android and Windows devices.

\textbf{Takeaways. }Before proceeding to the modeling part, we inspected all three generated patterns (TWS, RCA and SCU) for diversification of information. In terms of information, we find that TWS is the richest of the three patterns, followed by RCA, which is then followed by SCU (see Section 4 in the \href{https://drive.google.com/drive/folders/1CvsAuMG15L2Xk71XXkqLDT0DWlxHVr9l?usp=sharing}{supplementary document}). In the next section, we investigate our findings in light of the premise that the TWS collects more diverse information by predicting socio-economic features.


\section{Predicting socio-economic indicators}\label{sec:Exp}
Here, we build machine learning models using the patterns (week signature, RCA index, and SCU index) we explored in the previous sections to predict the socio-economic features of IRIS in France.

\subsection{Features Used For Learning}
To illustrate the predictive power of the different feature sets reported earlier in this paper, we define a series of models, each with a distinct feature set corresponding to mobile app usage patterns. We focus on three types of features:
    
\noindent 1) \textbf{IRIS weekly signature: }In Section \ref{subsec:eda_weekly_signature}, we observed that the week signature of the apps category could distinguish the IRIS based on their socio-economic features.
    
\noindent 2) \textbf{IRIS Revealed Comparative Advantage (RCA) utilization: }In Section \ref{subsec:eda_rca_index}, we found that the RCA index patterns of various app categories are more or less utilized based on socio-economic features. Multiple apps' distinctive nature can help predict an IRIS's socio-economic features.
    
\noindent 3) \textbf{IRIS standardized cumulative utilization: }In Section \ref{subsec:eda_scu_values}, we observed SCU index for app categories could be used to predict socio-economic features.

\subsection{Experimental Setup}
We experiment with several regression models, including Linear Regression, Ridge Regression, and CatBoost. We find that CatBoost models are the most effective for our task, and we thus report results from only those models. Because of the regression task, we take care of the trade-off between bias and variance of the model. During experimentation, we split our dataset into 80:20 train and test sets, ensuring that our CatBoost models are not overfitting our training data. To achieve this, we used a 5-fold cross-validation on training data. Finally, we report the R-squared value on test data.

\subsection{Results with Explainability}\label{subsec:results&Explainability}
We subdivide our results into three parts based on economic status, population information, and education information (please see Section \ref{sec:SocioEconomic} for detail). 

\noindent \textbf{1) Economic status: }Table \ref{table:results} shows the R-squared value for the prediction of our models. We used four variations of predictive features to predict Poverty, Median income, and the Gini index. The variations of predictive features include (1) \textit{Cumulative: }that represents SCU values, (2) \textit{RCA: }that includes RCA index values, (3) \textit{TWS: }that has typical week signature on hourly basis, and (4) \textit{All: }which contains all the three mentioned features combined, i.e., SCU values, RCA index, and TWS.

\begin{table}
\centering
\begin{tabular}{lllll}
\hline
Socio-Economic Features & Cumulative & RCA   & TWS & All    \\\hline
\textit{Economic status}      & & &\\
Poverty                   & 0.306      & 0.366 & 0.458     & \textbf{0.482}  \\
Median Income             & 0.551      & 0.592 & 0.631     & \textbf{0.659}  \\
Gini Index                & 0.569      & 0.592 & 0.623     & \textbf{0.642}  \\
\hline
\textit{Education information}      & & &\\
No diploma                & 0.380      & 0.392 & 0.450     & \textbf{0.456}  \\
Baccalauréat              & 0.442      & 0.454 & 0.508     & \textbf{0.516}  \\
BEPC + CAPBEP             & 0.381      & 0.398 & 0.457     & \textbf{0.464}  \\
College                   & 0.538      & 0.548 & 0.596     & \textbf{0.604}  \\\hline
\textit{Population information}      & & &\\
Total population          & 0.457      & 0.470 & 0.520     & \textbf{0.527}  \\
Population 0 to 14 yrs    & 0.464      & 0.480 & 0.538     & \textbf{0.543}  \\
Population 15 to 29 yrs   & 0.529      & 0.541 & 0.591     & \textbf{0.596}  \\
Population 30 to 44 yrs   & 0.483      & 0.496 & 0.553     & \textbf{0.560}  \\
Population 45 to 59 yrs   & 0.449      & 0.461 & 0.516     & \textbf{0.522}  \\
Population 60 to 74 yrs   & 0.393      & 0.406 & 0.462     & \textbf{0.469}  \\
Population 75+ yrs        & 0.368      & 0.378 & 0.428     & \textbf{0.435}  \\
Immigrants                & 0.596      & 0.605 & 0.654     & \textbf{0.664}  \\
Farmer                    & 0.246      & 0.256 & 0.322     & \textbf{0.328}  \\
Craftsman, Trader         & 0.363      & 0.372 & 0.407     & \textbf{0.417}  \\
Manager, high profession  & 0.544      & 0.555 & 0.593     & \textbf{0.602}  \\
Intermediate profession   & 0.473      & 0.484 & 0.537     & \textbf{0.543}  \\
Employees                 & 0.437      & 0.449 & 0.498     & \textbf{0.504}  \\
Worker                    & 0.354      & 0.369 & 0.432     & \textbf{0.437}  \\
Retired                   & 0.359      & 0.371 & 0.425     & \textbf{0.433}  \\
Other w/o profession      & 0.485      & 0.494 & 0.536     & \textbf{0.540}  \\
Male population           & 0.459      & 0.472 & 0.525     & \textbf{0.532}  \\
Female population         & 0.467      & 0.478 & 0.529     & \textbf{0.535} \\\hline
\end{tabular}
\caption{R-squared scores using \textit{Cumulative}, \textit{RCA}, \textit{TWS}, and \textit{All} predictive features for the socio-economic features. Best score is highlighted using bold text.}\label{table:results}
\end{table}

\begin{figure*}[ht!]
  \centering
  \subfloat[Poverty]{\label{fig:shap_poverty}\includegraphics[width=0.63\columnwidth]{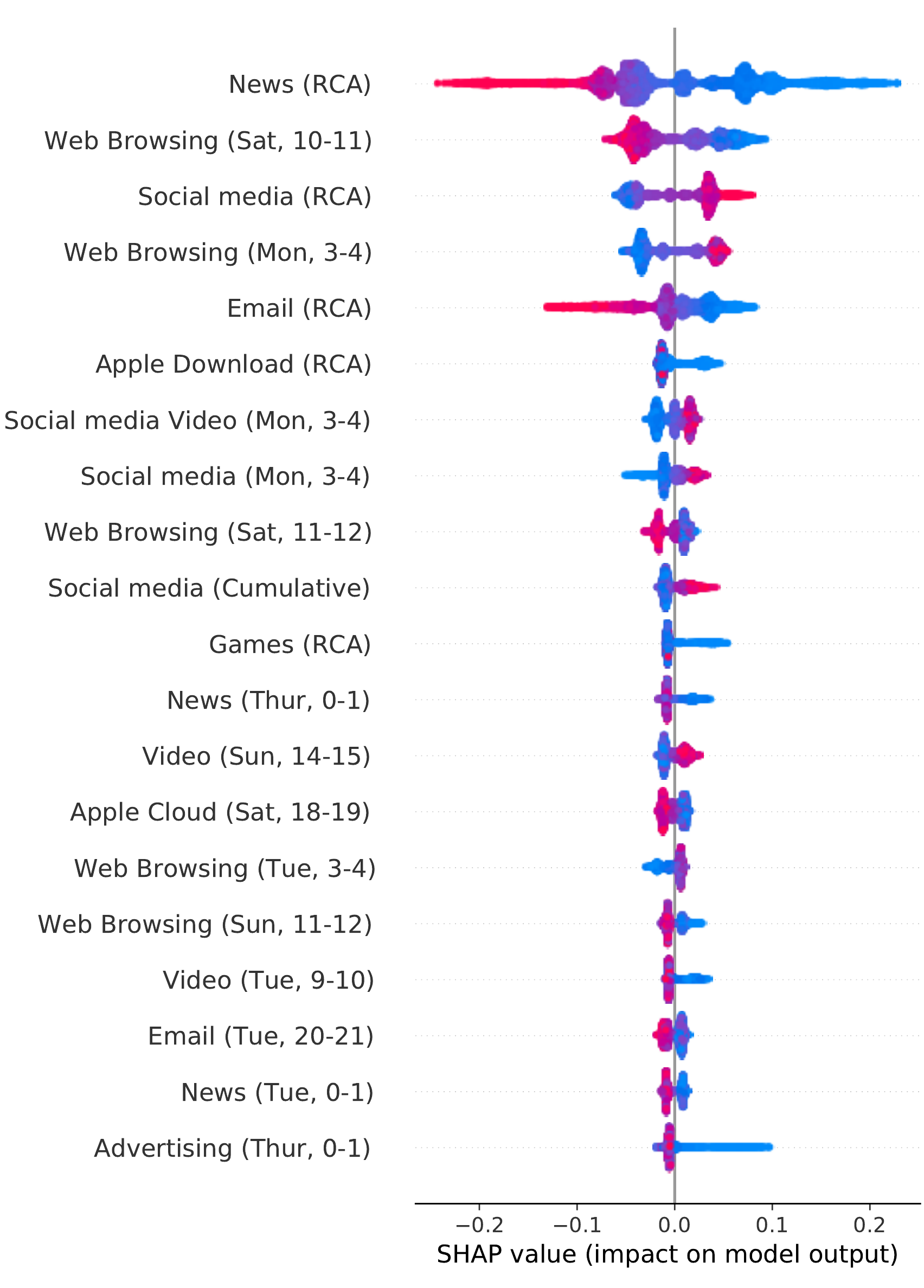}}\hspace{1mm}
  \subfloat[Median Income]{\label{fig:shap_median_income}\includegraphics[width=0.63\columnwidth]{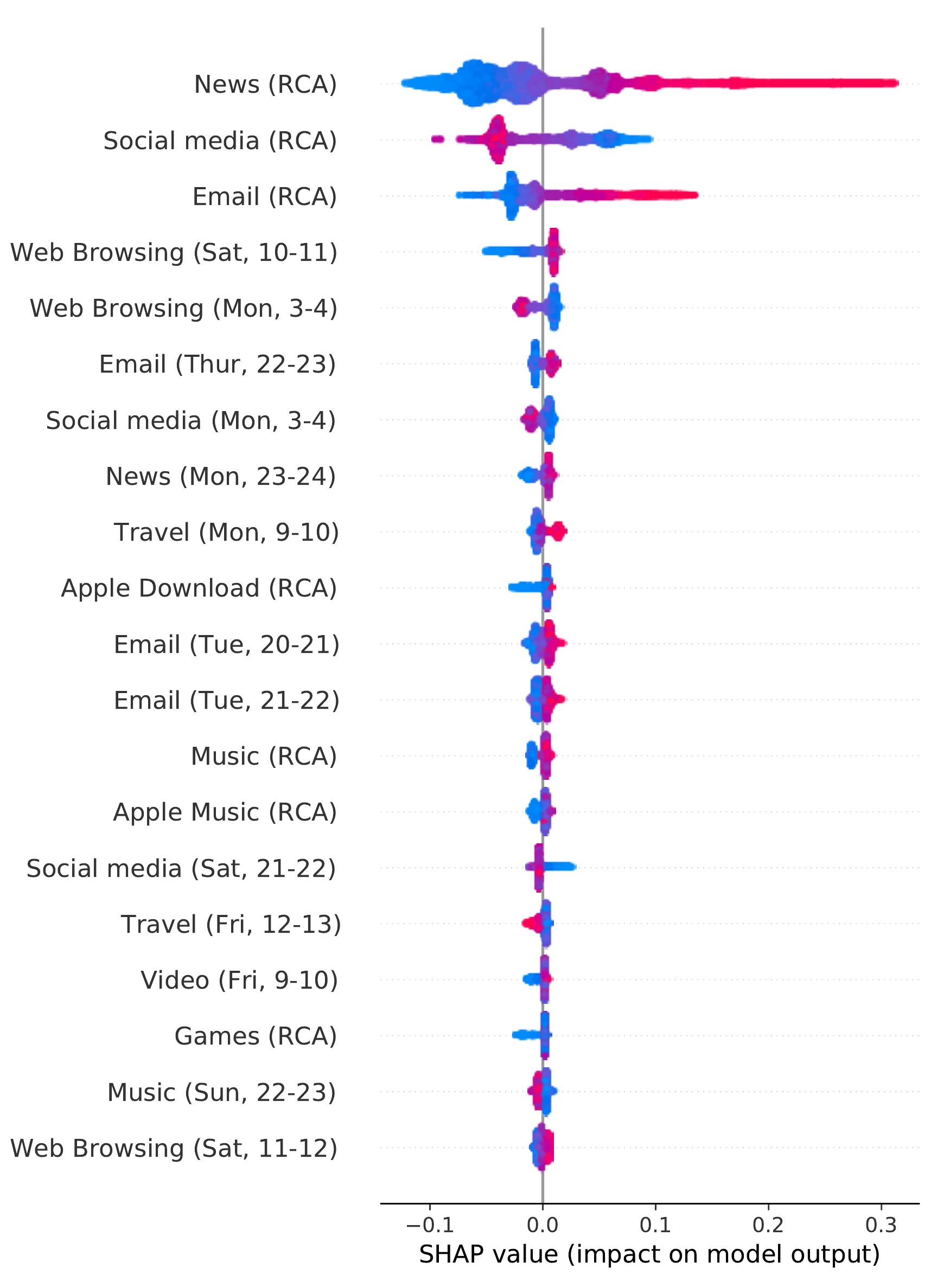}}\hspace{1mm}
  \subfloat[Gini Index]{\label{fig:shap_gini_index}\includegraphics[width=0.74\columnwidth]{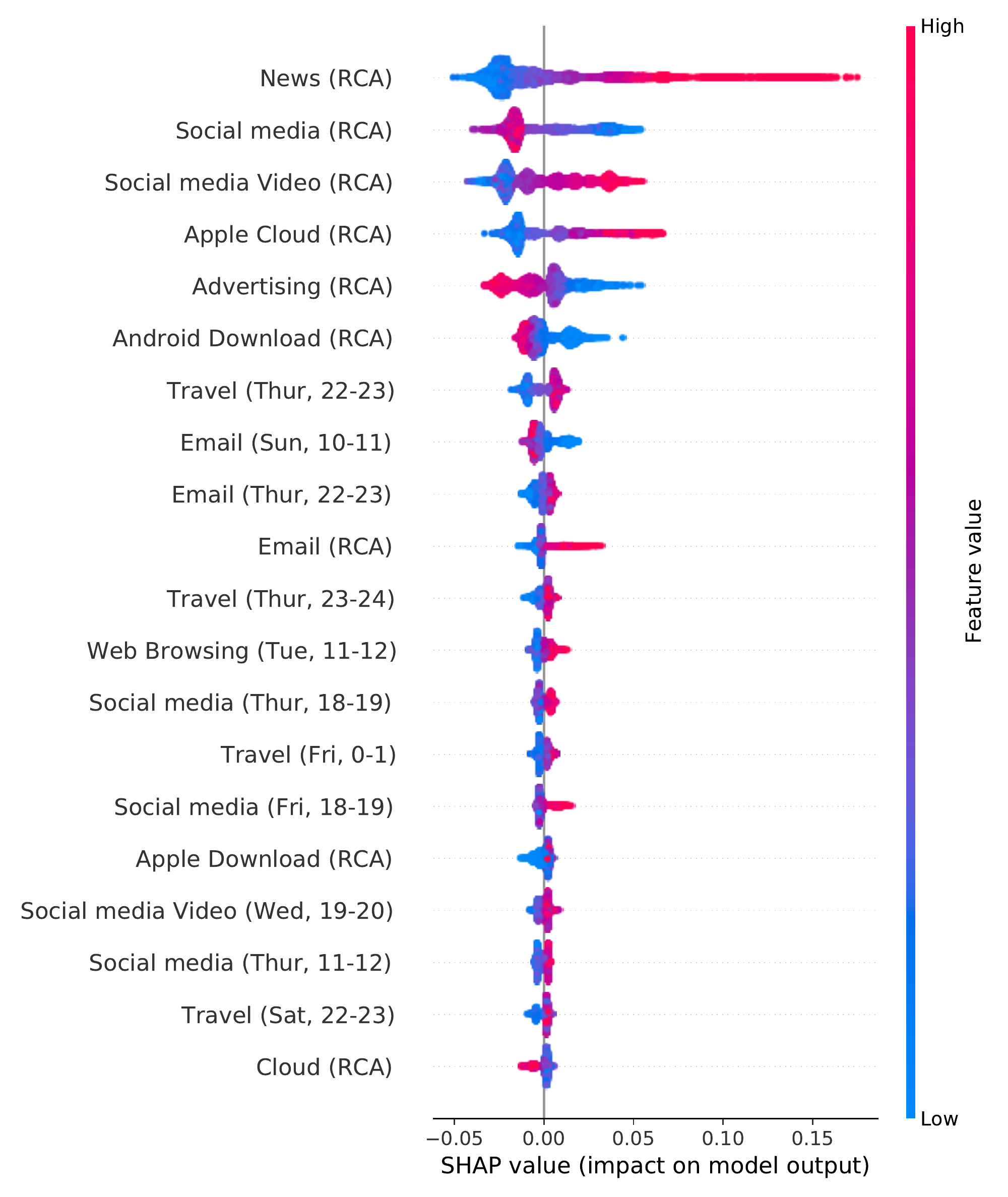}}
  \caption{SHAP plot to determine feature importance for Economic status. The SHAP values and feature names are represented by the x- and y-axes, respectively. Each data point represents a single instance. The \textcolor{red}{red} color represents a higher value for the feature than its average value, while the \textcolor{blue}{blue} color denotes a lower value. \textcolor{red}{Red} values on the right side of the x-axis indicate a positive impact on the prediction and vice versa.  The features are listed in order of decreasing importance (best seen in color).}
\label{fig:economic_condition_expl}
\end{figure*}

We find the most predictive features to be the week signature. Strikingly, mobile app usage at different times of day is significantly more predictive than the RCA index and cumulative usage patterns. With a model trained on all available features, we achieve a score of 0.48, 0.66, and 0.64 for Poverty, Median income, and Gini index, respectively. The results indicate that the mobile app usage patterns, particularly the TWS, have the capability to predict the reported socio-economic features. 

Furthermore, to understand the features' importance for the best model with \textit{All} predictive features, we use SHAP (Shapley additive explanations) values~\cite{lundberg2017unified}. A summary plot of SHAP values for top-20 features is shown in Figure \ref{fig:economic_condition_expl}. Each point represents an IRIS, thus for each feature, the figure shows the distribution of SHAP values across the dataset. The wider distribution indicates a larger absolute impact of the feature in the prediction, while the color of each point encodes the feature value (low or high). A feature with high values corresponding to negative SHAP values (to the left) is negatively correlated with economic status. For example, RCA index for \textit{News} app category or \textit{News (RCA)} is negatively correlated with \textit{Poverty} (see Figure \ref{fig:shap_median_income} and \ref{fig:shap_gini_index}). Conversely, features with high values corresponding to positive SHAP values (to the right), are positively correlate with economic status. For example, \textit{News (RCA)} is positively correlated with \textit{Median income} and \textit{Gini index} (see Figure \ref{fig:shap_median_income} and \ref{fig:shap_gini_index}).

The most important feature for predicting economic status are News, Web Browsing, Social media, Email, Social media video, etc. We observe that \textit{News (RCA)} consumption is more in IRIS with high median income and a high Gini index. On the other hand, \textit{News (RCA)} consumption is less in IRIS with high Poverty. That means News is read more by high-income individuals. Additionally, RCA index for \textit{Social media} app category or \textit{Social media (RCA)} usage pattern is completely opposite compared to \textit{News (RCA)}. That indicates that high-income people do not frequently use social media. Similarly, we find that \textit{Web Browsing} on Saturday from 10 to 12 is positively correlated with \textit{Median income} and negatively correlated with \textit{Poverty}. Conversely, we find that \textit{Web Browsing} on Monday from 3 to 4 is negatively correlated with \textit{Median income} and positively correlated with \textit{Poverty}. This implies that high-income individuals avoid browsing the internet on Monday late afternoon. Based on the conclusions above, it is worth to highlight this helpful property of TWS that allows identifying relationships over time between mobile app consumption and socio-economic features.

\textbf{Takeaways.} Social media applications are utilized less by the high-income group than by the low-income group. On the other hand, the high-income group frequently uses News and Productivity apps.

\noindent \textbf{2) Education: }Table \ref{table:results} reports the R-squared values for the CatBoost models for four education classes: No diploma, Compulsory Schooling (Baccalauréat), High School (BEPC and CAPBEP), and College. Again, we find the most predictive features to be the weekly signature. That means mobile app usage at different times of the day is significant. With a model trained on all available features, we achieve a score of 0.46, 0.52, 0.46, 0.60 for No diploma, Compulsory Schooling (Baccalauréat), High School (BEPC and CAPBEP), and College, respectively. 

\textbf{Takeaways. }Due to space limitation, the SHAP plots for education classes, are in the \href{https://drive.google.com/drive/folders/1CvsAuMG15L2Xk71XXkqLDT0DWlxHVr9l?usp=sharing}{supplementary document} (refer Figure S14). From the plots, we find that the consumption of social media videos, social media, music, and travel applications play a vital role in predicting the education status in the IRIS. We also observe that people without a diploma, compulsory education, and high school have more distinguished consumption of music and social media than those with a college education.

\noindent \textbf{3) Population: }Table \ref{table:results} report the R-squared values for the CatBoost models for all population features. We achieve scores between 0.44 (age 75+) to 0.66 (Immigrants). Our model predicted the immigrant population and people ages 15 to 29 years more accurately. Similarly, we achieve scores for profession information between 0.33 (Farmers) to 0.60 (Managers, high profession). The model performed better in predicting high profession, Intermediate profession, and Others with high scores.

\textbf{Takeaways. }The SHAP plots for population information are in the \href{https://drive.google.com/drive/folders/1CvsAuMG15L2Xk71XXkqLDT0DWlxHVr9l?usp=sharing}{supplementary document} (refer Figure S15). We observe that most of the features among the top 20 are from TWS. This further supports the relevance of TWS. We find that social media video is highly consumed by people aged between 15 to 45 years. The immigrant population, primarily young and working in positions with unprofessional activities, are not frequent targets for the advertisement. Additionally, high profession individuals, such as, Managers like to hear music after working hours.

\subsection{Relevance of the Findings}\label{subsec:relevance}
To show that the extracted patterns (TWS, RCA, and SCU) from mobile applications are better/comparable than census information for predicting socio-economic features, we further experimented using census data (Total pop, age 0-14, 15-29, 30-44, 45-59, 60-74, and 75+), and All (Signature, RCA, and Cumulative) to predict three socio-economic features: Median income, Gini index, and College education. Table \ref{table:results_comp} reports the R-squared values for the CatBoost models. We find that the model using \textit{All} predictive features achieves scores that are at least as good as those obtained with census data. 

\begin{table}
\centering
\begin{tabular}{lll}
\hline
Socio-Economic Features & Census & All \\\hline
Median income                   & 0.50      & \textbf{0.66}  \\
Gini Index            & 0.38      & \textbf{0.64}  \\
College                & \textbf{0.61}      & \textbf{0.61}  \\\hline
\end{tabular}
\caption{R-squared scores for CatBoost model using \textit{census}, and \textit{All (Cumulative, RCA, and TWS patterns)} data for predicting Median income, Gini index, and College education. Best score is highlighted using bold text.}\label{table:results_comp}
\end{table}

Please note that, for the sake of completeness, we performed a comparability analysis between the mobile applications usage patterns and census information for predicting three socio-economic features: Median income, Gini index, and College education. However, the use of population information for predicting socio-economic features is not aligned with the purpose of this study. That is, this study tries to 
provide an alternative to expensive, and time-consuming census tasks to collect socio-economic information by mobile usage to predict socio-economic features.


\section{Conclusion}\label{sec:conclusion}
This work proposes a large-scale, quantitative, and predictive study of the relationship between mobile app usage and socio-economic features. We analyzed nationwide data from the leading mobile operators in France and extracted three patterns: TWS, RCA, and SCU. We find that TWS has richer information diversification than RCA and SCU. The best model using all three patterns achieved an R-squared score up to 0.66, thus concluding our study goal that mobile application usage can predict a nation's socio-economic conditions.

\noindent \textbf{Impact: }Most countries conduct a census once every ten years and utilize the same data to drive policies prior to the next census~\cite{un2017principles,pogo}. To overcome the limitation of census, this study makes a case for an inexpensive, privacy-preserving, real-time and scalable method to understand the latest socio-economic conditions and, by extension, poverty, unemployment, literacy, or economic progress in our societies through mobile phone data (application usage).

\noindent \textbf{Limitations and Future scope: }Obviously, the outcomes of this study are confined to France and its specific geographical granularity, IRIS. Studies of other countries' call data records or mobile application usage will need to define the geographical granularity required for this study. Another limitation of this study when duplicating it in other countries is getting socio-economic datasets for the desired geographical granularity. 
This study's findings suggest numerous possibilities for future directions, including user temporal network analysis to understand emerging and evolving network patterns. Another future direction is to investigate alternate sources of information that can be used in conjunction with or independent of mobile data to examine socio-economic conditions. 


\section*{Acknowledgment}
This research is supported by EU H2020 program under the SoBigData++ project (grant agreement No. 871042), the French ANR research projects DISCRET (grant number ANR-19-FLJO-0002-01) and PROMENADE (grant number ANR-18-CE22-0008). 





%

\begin{IEEEbiography}[{\includegraphics[width=1in,height=1.25in,clip,keepaspectratio]{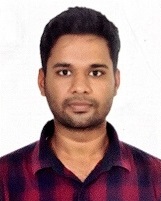}}]{Rahul Goel} is a PhD scholar at the University of Tartu, Estonia. He has BTech in IT and MTech from NIT Delhi, India in Computer Science and Engineering (Analytics). He also worked as a data scientist and ASE in Impact Big Data Analytics and Centre for Railway Information System respectively. His research interests include social media study, network analysis, and natural language processing using data science, machine learning, and deep learning.
\end{IEEEbiography}

\begin{IEEEbiography}[{\includegraphics[width=1in,height=1.25in,clip,keepaspectratio]{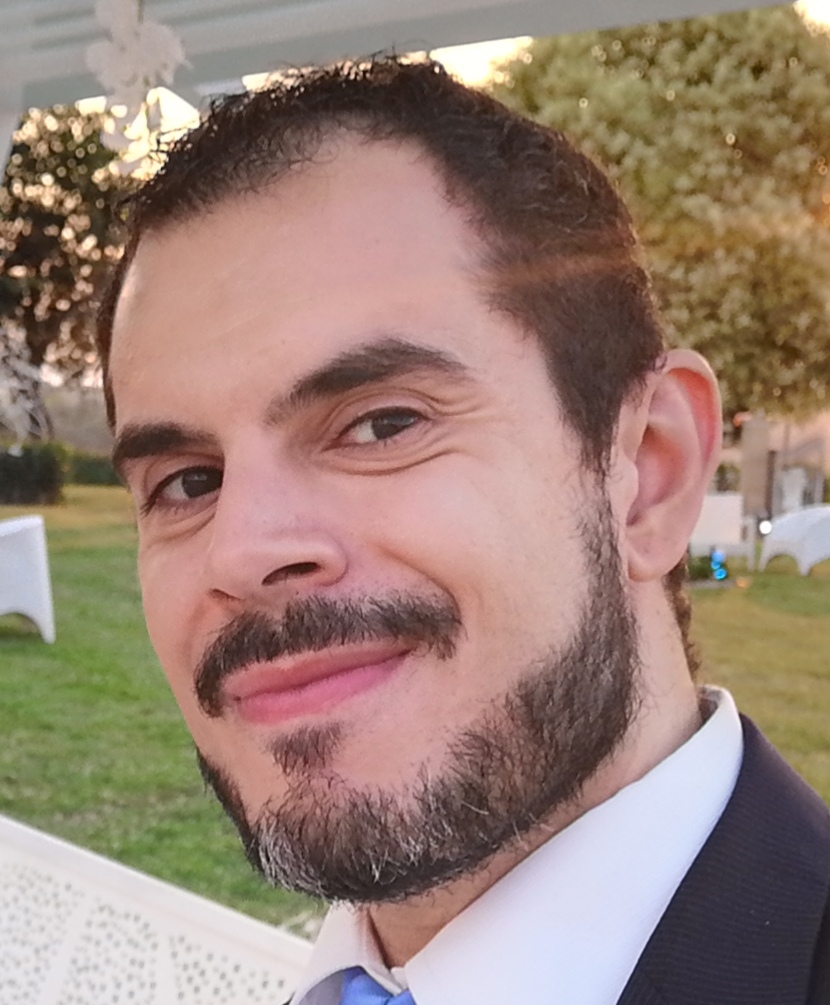}}]{Angelo Furno} is a researcher at the ENTPE, University of Lyon, and the University Gustave Eiffel, France, where he has been a member of LICIT-ECO7 laboratory since 2016. From 2014 to 2016, he was a postdoctoral researcher with INRIA at the CITI laboratory of INSA-Lyon, France. He obtained his Ph.D. in computer science from the University of Sannio, Italy. His research interests include urban computing, smart mobility, big data, machine learning, mobile phone data analytics, and distributed computing. He is the author of more than 50 contributions in leading international journals and conferences in related fields.
\end{IEEEbiography}

\begin{IEEEbiography}[{\includegraphics[width=1in,height=1.25in,clip,keepaspectratio]{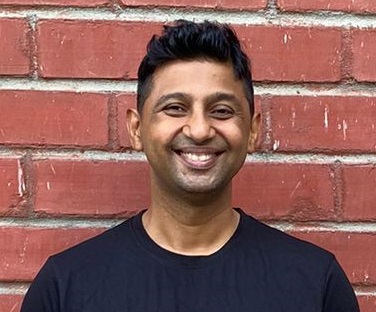}}]{Rajesh Sharma} is presently working as associate professor and leads the computational social science group at the Institute of Computer Science at the University of Tartu, Estonia, since January 2021. 

Rajesh joined the University of Tartu in August 2017 and worked as a senior researcher (equivalent to Associate Professor) till December 2020. From Jan 2014 to July 2017, he has held Research Fellow and Postdoc positions at the University of Bristol, Queen's University, Belfast, UK and the University of Bologna, Italy. Prior to that, he completed his PhD from Nanyang Technological University, Singapore, in December 2013. He has also worked in the IT industry for about 2.5 years after completing his Master's from the Indian Institute of Technology (IIT), Roorkee, India. Rajesh's research interests lie in understanding users' behavior, especially using social media data. His group often applies techniques from AI, NLP, and most importantly, network science/social network analysis.
\end{IEEEbiography}




\end{document}